\documentstyle[emulateapj,graphicx]{article}
\def\gsim{\;\rlap{\lower 2.5pt
 \hbox{$\sim$}}\raise 1.5pt\hbox{$>$}\;}
\def\lsim{\;\rlap{\lower 2.5pt
   \hbox{$\sim$}}\raise 1.5pt\hbox{$<$}\;}
\newcommand{\beq}{\begin{equation}}
\newcommand{\eeq}{\end{equation}}
\def\myputfigure#1#2#3#4#5%
{\hskip0.03\textwidth\vskip#5pt
\makebox[0pt]{\hskip#2in
\includegraphics[width=#3\textwidth]{#1}}\vskip#4pt\hfill}


\lefthead{Haiman, Spergel \& Turner}
\righthead{Mid-Infrared Counts of High-Redshift Galaxies}

\begin{document}
\title{Predictions for the Counts of Faint, High-Redshift Galaxies in the Mid-Infrared}
\author{Zolt\'an Haiman\altaffilmark{1}, David N. Spergel, and Edwin L. Turner}
\affil{Princeton University Observatory, Princeton, NJ 08544, USA\\ 
zoltan,dns,elt@astro.princeton.edu}
\altaffiltext{1}{Hubble Fellow}

\vspace{\baselineskip}
\begin{abstract}
Deep mid--infrared (MIR) observations could reveal a population of
faint, high--redshift ($z>3$) dusty starburst galaxies that are the
progenitors of present--day spheroids or bulges, and are beyond the
reach of current instruments. 
We utilize a semi--analytic galaxy formation scheme to find an extreme
model for the MIR galaxy counts, designed to maximize the number of
detectable sources down to a flux level of a few nJy.  The model
incorporates the formation of heavily dust--enshrouded stellar
populations at high redshift, and is consistent with existing
observations, including faint counts at 1.6$\mu$m in the NICMOS Hubble
Deep Field, and the upper limit on the extragalactic MIR background
from TeV gamma rays.
Our models predict upto $\sim$0.5 galaxies/arcsec$^2$ at the threshold
of 100 nJy at 6$\mu$m, with a comparable or larger surface density at
longer MIR wavelengths.  We conclude that a significant new population
of high--redshift galaxies could be detected by the {\it Space
Infrared Telescope Facility (SIRTF)} and {\it Next Generation Space
Telescope (NGST)}. Such a population would constitute background noise
for the {\it Terrestrial Planet Finder (TPF)}, and could necessitate
repeat observations: every {\it TPF} resolution element have a $\sim 10\%$
chance of being contaminated by a background galaxy.

\end{abstract}

\section{Introduction}
\label{sec:introduction}

The past few years have seen significant progress in probing the
ultra--high redshift universe, with both galaxies (Dey et al. 1998;
Weymann et al. 1998; Spinrad et al. 1998; Hu et al. 1999) and quasars
(Fan et al. 1999, 2000, 2001; Zheng et al. 2000; Stern et al. 2000)
being discovered in increasing numbers well beyond redshift $z=5$.  In
hierarchical structure formation scenarios in cold dark matter (CDM)
cosmologies, the first baryonic objects appear at still higher
redshifts: at $z\approx 20-30$, when the first high--$\sigma$ peaks
collapse near the Jeans scale of $\sim 10^5~{\rm M_\odot}$ (Haiman,
Thoul \& Loeb 1996; see Barkana \& Loeb 2001 for a recent review).
Radiative cooling is efficient in the dense gas that has collapsed on
these scales, and in principle, it can facilitate efficient
star--formation.  Indeed, significant activity must have taken places
at high redshifts, in order to reionize the intergalactic medium (IGM)
by $z\gsim 6$, and enrich it with metals by $z\gsim 4$.

The deepest detections of galaxies and quasars to date have been
obtained at optical or near infrared (NIR) wavelengths, where the
objects were identified in broad--band filters by their continuum, or
in narrow--band imaging observations by their Lyman--$\alpha$
emission.  The {\it Next Generation Space Telescope (NGST)} will be
able to extend these observations to $\gsim 32$ mag in the $1-5\mu$m
wavelength range, and detect mini--galaxies and mini--quasars at
redshifts $z\gsim 10$.  The expected number of faint sources in
future, deep NIR observations have been studied extensively in the
context of hierarchical structure formation, using simple
semi--analytic models.  Haiman \& Loeb (1997; 1998) showed that if
halos collapsing at high redshifts have reasonable star (or quasar
black hole, BH) formation efficiencies, they can then be detected in
the NIR continuum in great numbers, with surface densities possibly
reaching $\sim1000$ sources per arcmin$^{-2}$.  Similarly large
numbers of high--redshift objects could be detected through
optical/NIR narrow band filters or spectroscopic imaging.  The counts
have been computed and found to be potentially significant for
Ly$\alpha$ emission originating either from a usual stellar population
(Haiman \& Spaans 1999) or from the release of gravitational binding
energy (Haiman, Spaans \& Quataert 2000).  In addition, recombination
lines of helium fall into the optical/NIR, allowing the detection of
high--redshift sources, provided they have sufficiently hard spectra
(Tumlinson \& Shull 2001; Oh \& Haiman 2001).

Observations at nearby redshifts have revealed that spheroid systems -- the
bulges of disk galaxies, as well as dwarf spheroidal galaxies -- have
exceedingly old stellar populations (see, e.g. Binney \& Tremaine 1987).  It is
natural to assume that these objects formed at high redshifts.  At the epoch
when the halos harboring these objects first assembled, gas supply was likely
plentiful, resulting in high star formation rates.  In analogy with local
starburst galaxies, these high--redshift bursts of star formation were likely
heavily dust enshrouded, with unusually red spectra enhancing fluxes at longer
wavelengths.  In this paper, our goal is to quantify this scenario, and to
predict the counts of faint, high--redshift galaxies at mid--IR (MIR)
wavelengths.

Semi--analytic galaxy formation models, originally applied at optical
wavelengths (Kauffmann \& White 1993) have recently been extended to the
far--infrared (FIR), and all the way to sub-mm range.  The key to such
extensions is the availability of template spectra that incorporate the
absorption and re--emission of starlight by dust.  Dusty galaxy models have
successfully matched spectra of known starburst galaxies (Gordon et al. 1997),
as well as a broader range of galaxy types (Silva et al. 1998; Devriendt et
al. 1999).  When combined with hierarchical galaxy--formation schemes, such
spectral models have also successfully reproduced the existing IR/sub--mm
luminosity functions (Guiderdoni et al. 1998; Silva et al. 1999), and have been
used to investigate several aspects of IR galaxies, such as the faint--end
slope of their luminosity function, and the abundance of ultra--luminous
infrared galaxies (ULIRGs, Devriendt \& Guiderdoni 2000).

In the present paper, we consider the number counts of faint, high redshift
sources at MIR wavelengths, using similar semi--analytic models.  The main
difference between the present paper and previous studies is that we
extrapolate the models down to a very faint flux level.  Our study is motivated
primarily by the forthcoming instruments {\it NGST}, the {\it Space Infrared
Telescope Facility (SIRTF)}, and the {\it Terrestrial Planet Finder (TPF)}. It
is likely that {\it NGST} will have very deep ($\sim 100$nJy) imaging
capability in the MIR out to $\lambda\sim30\mu$m (Serabyn et al. 1999).  In
very long exposure $10^6$s observations, {\it SIRTF} could reach similar
limiting fluxes; while $\sim 100$nJy is also the target flux level for the IR
version of {\it TPF} to discover Earth--like planets at a distance of 10pc.

Observations in the MIR have only been possible in a few narrow bands
from the ground, and the deepest existing surveys from space, i.e. by
the {\it Infrared Space Observatory (ISO)}, are still relatively
shallow, achieving completeness only down to $\sim 0.1$mJy (see,
e.g. Franceschini 2000 for a review).  We use the models to obtain
counts to the much fainter flux levels of $\sim 1$nJy. We emphasize
that this is a very significant extrapolation from current data, by
several orders of magnitude. Such extrapolations are inevitably
uncertain. In particular, we do not attempt here to present a ``most
likely'' model.  Instead, the goal of the present paper is to produce
``extreme'' models that {\it maximize} the MIR counts at $\sim$100
nJy, but are (1) consistent with all existing observations, and (2)
are not obviously physically unrealistic.  These predictions will
serve as a guide to the most optimistic scenario for detecting
ultra--faint galaxies with {\it NGST}, {\it SIRTF}, and {\it TPF}. In
addition, these calculations will be useful to assess whether these
observations may reach the MIR confusion limit.

The rest of this paper is organized as follows. In \S\ref{sec:model},
we describe the ingredients of our modeling, including the presence of
dust, and discuss the relevant observational constraints. In
\S\ref{sec:results}, we present the counts at different MIR wavelengths,
describe the properties of the faint sources, such as typical masses
and redshift distributions, and discuss the confusion limit.  Finally,
in \S\ref{sec:conclusions}, we summarize our conclusions and the
implications of this work. Throughout this paper, we assume a flat
$\Lambda$CDM cosmology with the parameters $(\Omega_{\rm
m},\Omega_{\Lambda},\Omega_{\rm b}h^{2},h,\sigma_{8
h^{-1}})=(0.3,0.7,0.019,0.7,1.0)$.

\section{Model Description}
\label{sec:model}

Our semi--analytical approach is a simplified version of the
Monte--Carlo models found in the literature of hierarchical galaxy
formation (Kauffmann \& White 1993, Lacey \& Cole 1993; for a more
recent application to high--redshift galaxies, see Haiman \& Loeb
1997, 1998; Wechsler et al. 2001).  Its three main aspects are (1) the
distribution of dark matter halos; (2) the template spectra and
light--curves of the stellar populations; and (3) the calibration of
the star--formation efficiency.  In this section, we describe our
treatment of each of these issues in turn.

\subsection{Dark Matter Halos}
\label{subsec:halos}

We assume that galaxies form in dark matter halos, whose abundance
$dN/dM(M,z)$ follows the standard Press \& Schechter (1974; hereafter
PS) mass function.  The cosmological power spectrum is computed from
the fitting formulae of Eisenstein \& Hu (1999), and we set
$\delta_c=1.68$ for the usual critical overdensity for collapse.  In
the extended PS formalism, it is also possible to compute the
distribution of ages for halos of a given mass and redshift. Here we
define the age of a halo to be the time elapsed since the halo first
acquired half of its mass, and follow equation 2.26 in Lacey \& Cole
(1993; hereafter LC) to obtain the halo age--distribution
$dp/dt(M,z)$.  We assume further that the age of the stellar
population in the halo equals the age of the halo.

It is important to emphasize that improvements have been made over the
PS mass function, taking into account the lack of spherical symmetry,
and that large--scale three--dimensional simulations have possibly
uncovered significant differences in the abundance of high--$\sigma$
objects (e.g., Sheth, Mo \& Tormen 2001; Jenkins et al. 2001).  Our
main motivation for the choosing the standard PS mass function is
``technical'': the semi--analytical derivation of $dp/dt$ is only
applicable for the mass function in the standard PS theory; at present
no analogous derivation exists for the improved mass functions.
However, we note that the typical objects we will be interested in
below have halo masses corresponding to $\lsim 3.5\sigma$ density
peaks. At these masses, the discrepancy between the PS mass function
and the simulations is within a factor of $\sim$ three, with the PS
formula under--predicting the abundance.  Hence, we expect that if we
were able to use the improved mass function (e.g. equation 9 in
Jenkins et al. 2001), at any given flux and at the highest redshifts,
the number of sources we predict could increase by upto a factor of
$\sim$three (or alternatively, we would predict the same number of
sources for a $\sim$three times lower star--formation efficiency).

\subsection{Template Spectra and the Effects of Dust}
\label{subsec:spectra}

Another ingredient of our model is the template spectrum emitted by
high--redshift galaxies.  Standard population synthesis models have
been successful in matching the optical/NIR spectra of observed nearby
galaxies (see Leitherer et al. 1999 for a review), and have often been
adopted in semi--analytical galaxy formation models. Closest to the
present context is the study of counts at high redshifts focusing on
these wavelengths (e.g. Haiman \& Loeb 1997; 1998).  However, as
mentioned above, the effects of dust can become conspicuous in the MIR
regime at wavelengths $\lambda\sim {\rm few}\,\mu$m, and the
importance of dust increases for still longer wavelengths.
Furthermore, studies of the spatial distribution and optical
properties of dust in nearby starburst galaxies have shown that models
with a simple ``foreground screen'' of ``normal'' dust (with a
cross--section similar to that of dust in the Milky Way or the
Magellanic Clouds) is inconsistent with the data (for a recent
example, see Gordon et al. 1997).  The inferences have been that the
stars and the dust must have different spatial distributions and
temporal evolutions.

The presence of dust therefore adds considerable complexity to
spectral modeling.  Several recent studies have addressed the problem
of dusty galaxy spectra (e.g. Silva et al. 1998; Devriendt et
al. 1999; Charlot \& Fall 2000). The main typical features of
successful models are that the dust is concentrated in dense
star--forming clouds, and therefore it has a highly patchy
distribution compared to the overall stellar distribution.  In
addition, star--forming clouds have a finite lifetime, making the
effects of dust time--dependent. The properties of the grains
themselves have also been found to be important in the MIR. In
particular, models for the size--distribution and cross--section of
grains accurately describe the optical/UV properties of Galactic dust
(Draine \& Lee 1984), but an additional hot dust component, requiring
the presence of very small grains or PAH molecules, is needed to
reproduce the observed spectra in the range
$2\mu$m$\lsim\lambda\lsim30\mu$m (e.g., Puget et al. 1985).

\myputfigure{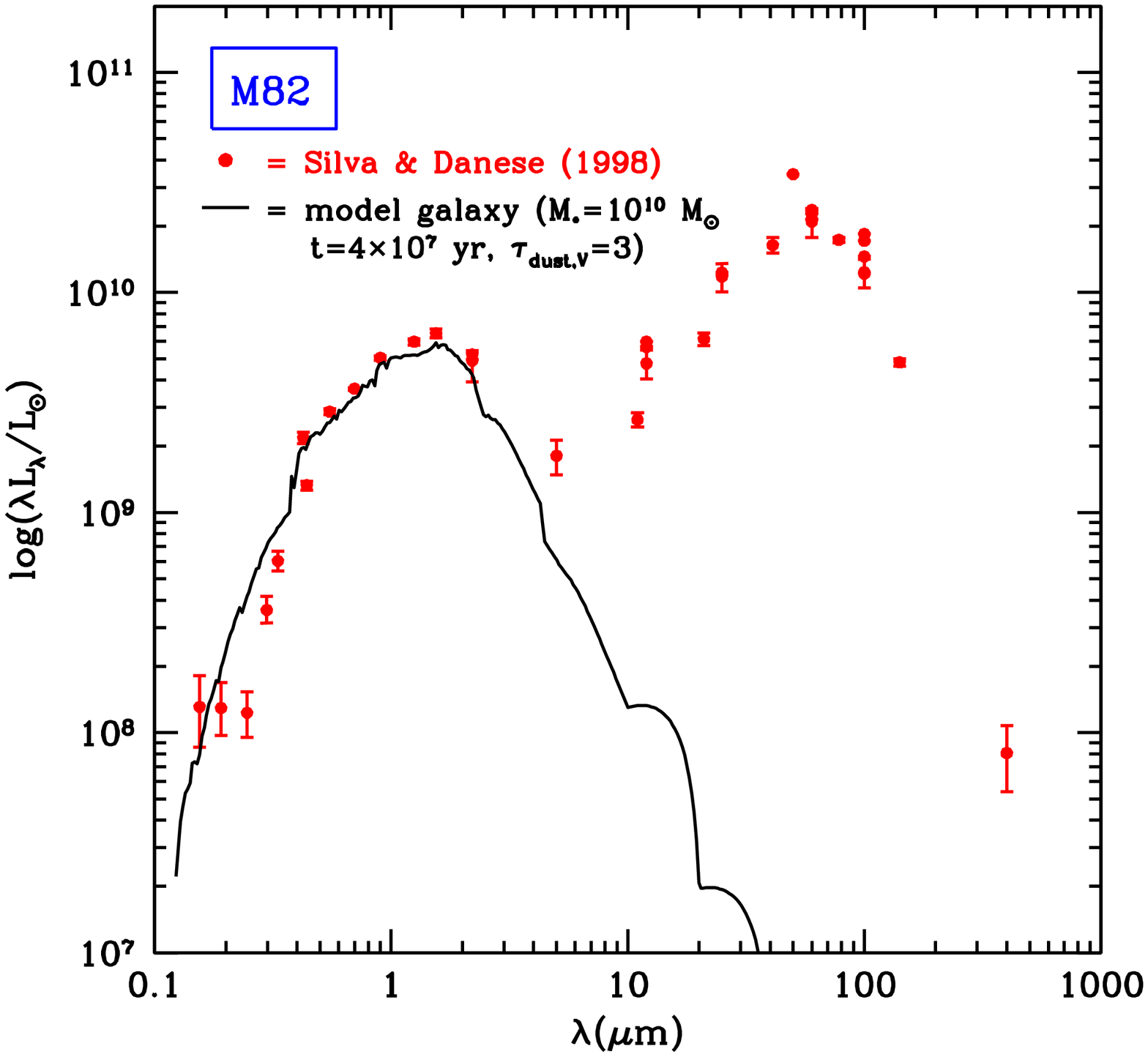}{3.2}{0.45}{-20}{-10} \figcaption{\label{fig:m82}
The spectrum of M82, a dusty starburst galaxy in the local universe,
is reproduced by a $9\times10^{9}~{\rm M_\odot}$ (stellar mass), and
$4\times10^7$yr old model galaxy upto $\sim5\mu$m.  The model does not
include dust emission, and under--predicts the flux at longer
wavelengths.}
\vspace{\baselineskip}

As emphasized by Charlot \& Fall (2000), as a result of these
complications, a simple foreground dust screen can not be assumed.
However, these authors derive a simple phenomenological recipe for the
effects of dust, which they find has an ``effective''
foreground--screen absorption curve proportional to $\lambda^{-0.7}$
(i.e. grey compared to Draine \& Lee).  In most of our calculations
below, we adopt this simple power--law absorption cross section.  The
overall normalization is still largely ad--hoc, and is likely to vary
from galaxy to galaxy.  However, here we chose the value such that
$\tau_{\rm dust}=3(\lambda/5500{\rm \AA})^{-0.7}$, i.e. we set the
opacity in the visual band to be $\sim 3$.  By assumption, this will
then represent the typical dust content of high--redshift starburst
galaxies.  This value is similar to that inferred for the local
starburst galaxy M82 (e.g. Silva et al. 1998; Devriendt et al. 1999).
In Figure~\ref{fig:m82}, we combine the dust--free stellar synthesis
models of Bruzual \& Charlot (2000) and a foreground screen of dust
with this opacity. The resulting spectrum provides an excellent fit to
the spectrum of M82 upto a wavelength of $\sim5\mu$m.  In this figure,
we have assumed a single burst of star formation with a stellar mass
of $9\times10^{9}~{\rm M_\odot}$, and an age of $4\times10^7$yr for
the stellar population.

\myputfigure{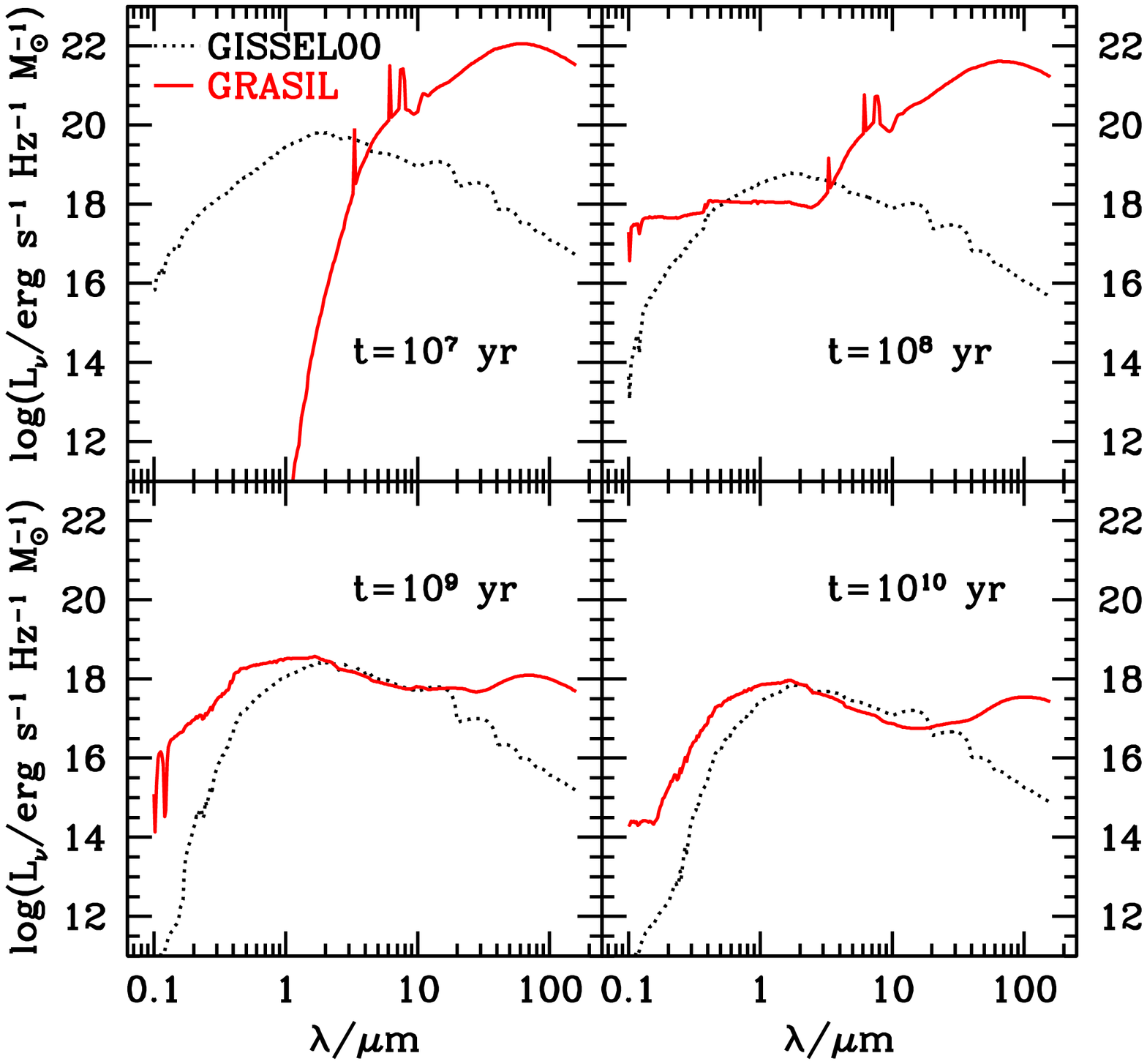}{3.5}{0.45}{-20}{-10} \figcaption{\label{fig:grasil}
Comparison of time--evolving template spectra adopted for galaxies,
under two different assumptions. The solid curves, labeled GISSEL00,
represent a model in which a dust--free population synthesis model is
combined with a foreground screen of dust with an effective absorption
law $\propto\lambda^{-0.7}$, as suggested by Charlot \& Fall (2000).
The dashed curves, labeled GRASIL, is based on the detailed model
spectra of a dusty starburst galaxy, using the computer code of Silva
et al. (1998).  The two models differ significantly at early stages
($t\lsim10^8$yr), but predict similar MIR spectra at late times.}
\vspace{\baselineskip}

An obvious shortcoming of this simple model is that it does not include dust
emission, and therefore under--predicts the flux at wavelengths
$\lambda\gsim5\mu$m.  However, we have verified explicitly that the total
amount of energy absorbed by the dust in the UV approximately equals the
observed FIR emission. Therefore at least energetically, our simplistic model
is viable, although it may prove difficult to produce the observed SN rate and
line emission (see Silva et al. 1998).  In order to make a more realistic
model, and in which we are also able to compute galaxy counts at longer
wavelengths, we utilized the publicly available program
GRASIL\footnote{Downloadable directly from http://grana.pd.astro.it}.  The
program computes the emergent time--dependent spectrum from a galaxy, under the
assumptions of patchy dust distribution (see Silva et al. 1998 for the
description of their model).  Here we used it to follow the evolution of the
spectrum after a single burst of dusty star formation, with the following
parameters (cf. Tables 1 and 2 in Silva et al. 1998): $M_{\rm
G}=1.8\times10^{10}~{\rm M_\odot}$ (total gas mass), $t_{\rm inf}=10^7~{\rm
yr}$ (gas infall timescale), $f_{\rm mc}=0.08$ (mass fraction of gas in
star--forming clouds that contribute most of the dust opacity), $r_{\rm
mc}=16$pc, $M_{\rm mc}=10^{6}~{\rm M_\odot}$, $t_0=5.7\times 10^7$yr (radius,
mass, and dispersion timescale of star--forming clouds), $r_c^*=0.15$kpc,
$r_c^c=0.15$kpc (core radii of star and gas distribution, both for a King
profile).  We then assumed that the normalization of the emitted flux scales
linearly with the total gas mass $M_{\rm G}$ (this implicitly requires that all
length--scales scale as $M_{\rm G}^{1/3}$).  This assumption allows us to
assign a time--dependent emission spectrum to a dark halo once its age and star
formation efficiency is specified. A more accurate interfacing of the template
spectra with the hierarchical galaxy formation models could include a
non--linear scaling between spectrum and galaxy mass (Silva et al. 1999).

For reference, we show the time--evolving template spectrum of a model
galaxy, per unit solar mass in stars, in Figure~\ref{fig:grasil}.  The
spectra are shown both under the assumption of an effective foreground
screen in a Bruzual--Charlot model (labeled GISSEL00), and using the
code GRASIL.  It is apparent that the two models predict quite
different MIR spectra at early times ($t\lsim 10^8$yr).  This stems
from the fact the in the GRASIL model, the initial starburst is
heavily dust--enshrouded. Nearly all of the starlight
$\lambda\lsim5\mu$m is absorbed by dust in molecular clouds, and
re--emitted at long wavelengths, until the star--forming clouds
disperse and the dust opacity is significantly reduced (at $t\gsim
t_0$).  At late times, the two models agree fairly well in the MIR
range ($5\mu$m$\lsim\lambda\lsim 30\mu$m), although at the longest
wavelengths, the GRASIL models still predict a larger flux.

\subsection{Calibration of Star--formation Efficiency, and Existing Constraints}
\label{subsec:constraints}

The final ingredient of our model is the calibration of the
star--formation efficiency in the dark halos.  Although this could
vary significantly from galaxy to galaxy, for simplicity we assume
here that all halos of a given velocity dispersion turn the same
amount of gas into stars.  There are several approaches to choosing a
calibration.  When fitting existing data, such as galaxy counts, then
the efficiencies can be chosen to be the best--fitting values
(e.g. Kauffmann \& White 1993).  In the models of Haiman \& Loeb
(1997, 1998) that extrapolate to high redshifts, the star--formation
efficiency was normalized based on the mean metallicity of the
high--redshift Ly$\alpha$ forest.

In the present work, our aim is to maximize high--redshift galaxy
counts.  Accordingly, we regard the overall normalization of the
starformation efficiency as a free parameter, and we set it to its
maximum allowed value based on existing constraints (see discussion
below).  We envision that the remnants of the high--redshift
starbursts can be identified with the spheroid components in local
galaxies. Accordingly, based on the Faber--Jackson relation, we adopt
the scaling $M_{\rm star}\propto \sigma_{\rm halo}^4$, where $M_{\rm
star}$ is the mass in turned into stars, and $\sigma_{\rm halo}$ is
the velocity dispersion of the host halo.  We then normalize the
models as follows:
\beq
\left(\frac{\sigma_{\rm halo}}{115~{\rm km~s^{-1}}}\right)^4=
\left(\frac{M_{\rm star}}{4.5\times10^{11}{\rm M_\odot}}\right)
\label{eq:fstar}
\eeq 
In addition, we postulate that no stars form in halos with velocity dispersions
less than $30~{\rm km~s^{-1}}$, because of the presence of the UV background
(see, e.g., Navarro \& Steinmetz 1997). Prior to reionization (which we here
assume to occur at redshift $z=10$), we lower this threshold to $11.7~{\rm
km~s^{-1}}$, corresponding to a virial temperature of $10^4$K, where this
cutoff is determined by the requirement of efficient cooling, rather than the
feedback from the UV background (e.g. Haiman, Abel \& Rees 2000). We also note
that equation~(\ref{eq:fstar}) corresponds to a $\sim3$ times higher
normalization of the Faber--Jackson relation than derived for the bulges of
local spiral galaxies (Whitmore, Kirshner \& Schechter 1979).

Using the standard relation between halo velocity dispersion and mass
(e.g. Navarro, Frenk \& White 1997), and assuming that the gas available for
star formation is $M_{\rm gas}=(\Omega_{\rm b}/\Omega_{\rm m})M_{\rm halo}$ the
stellar mass here corresponds nominally to $M_{\rm star}\sim 6-7\times M_{\rm
gas}$ (for the typical halos at each observed flux).  Hence our maximal model
is rather extreme, in that it assumes the formation of $\gsim6-7$ generations
of massive stars, formed in quick succession, to recycle the available gas into
stars 6--7 times (for reference, we note that the models in HL97 had the much
lower overall star formation efficiencies of $M_{\rm star}/M_{\rm gas}\sim
2-20\%$).  A stellar population with a Salpeter IMF would return only $\approx
30\%$ of its mass to the interstellar medium in $\sim 3\times10^8$ yr, and
would allow recycling of the gas only $\sim$twice. The requirement in our model
of 6--7 cycles could be achieved either with a flatter IMF (since massive stars
return essentially all their mass; and IMF slope of $\sim$1.8 instead of 2.35
is then required), or by postulating a larger gas reservoir for a system with a
given velocity version.  Note that significant metal enrichment, to solar
levels, implies that $\sim8$ generations of star--formation did indeed take
place in the Milky Way (e.g. Binney \& Tremaine 1987), and observed heavy
element abundances in galaxy clusters also favor significant enrichment at high
redshifts (Renzini 1997).  

\myputfigure{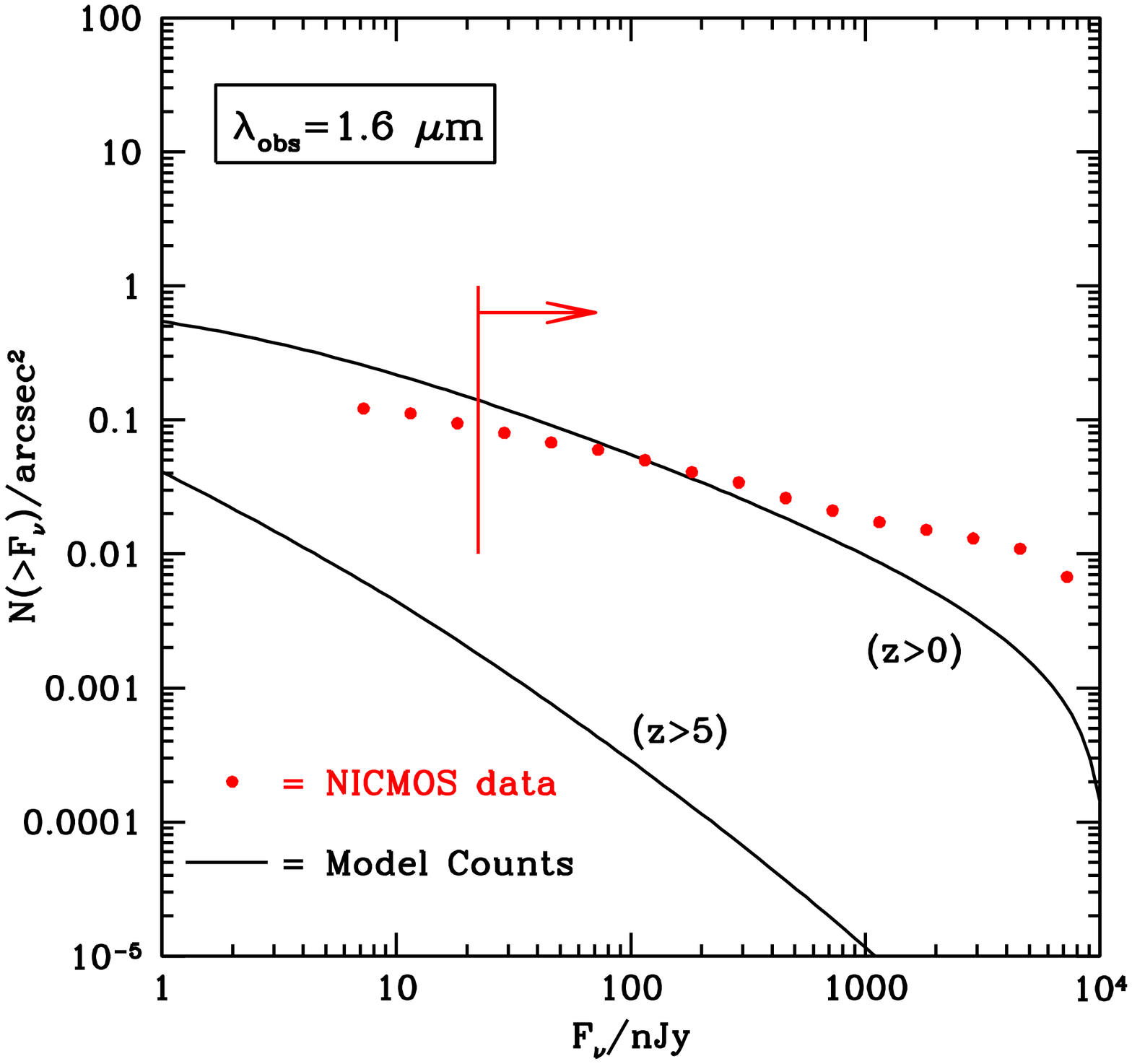}{3.2}{0.45}{-20}{-10} \figcaption{\label{fig:counts1}
Cumulative number counts of galaxies at 1.6$\mu$m brighter than a
given flux threshold $F_\nu$.  The upper solid curve shows all
galaxies, and the lower curve shows only those beyond redshift
$z=5$. The data--points are from NICMOS observations covering $1/8^{\rm
th}$ of the area of the Hubble Deep Field North (Thompson et
al. 1999).  Our models are consistent with these counts, owing to the
large dust opacity (and therefore red spectra) that we adopted.}
\vspace{\baselineskip}

Existing MIR counts (from {\it ISOCAM}) extend down only to about
$\sim0.1$mJy (Franceschini 2000; Franceschini et al. 1997; Clements et
al. 1999), and we extrapolate the models to several orders of
magnitude fainter flux levels.  Nevertheless, our normalization has to
be consistent with faint galaxy counts in the Hubble Deep Field (HDF)
in both optical and NIR bands.  In particular, we found that the most
constraining HDF data are the $1.6\mu$m galaxy counts in a NICMOS
follow--up observation of $\sim1/8^{\rm th}$ of the HDF area (Thompson
et al. 1999).  This deep survey has a 50\% completion limit near
28$^{\rm th}$ mag, and has detected a total of $\sim 300$ sources.  We
have found that models with the template spectra described in
\S~\ref{subsec:spectra} that are consistent with this abundance always
satisfy the limits from optical/UV counts in the Hubble Deep Field to
about the same depth.  In Figure~\ref{fig:counts1} we show the
$1.6\mu$m counts in our models using the GISSEL spectral models, and
with the normalization in equation~(\ref{eq:fstar}).  The upper curve
shows all sources, the lower curve shows only the sources beyond
redshift $z=5$, and the dots show the NICMOS data. The figure
explicitly demonstrates that our model is marginally consistent with
the NICMOS counts.

An integral constraint on the MIR counts can also be obtained from the
upper limit on the total cosmic infrared background energy density.
The latter limit derives from the TeV gamma ray spectrum of the blazar
Mrk 501, observed in its high state with HEGRA, yielding a stringent
limit on the optical depth to pair production at TeV energies (Stanev
\& Franceschini 1998; Dwek 2001).  The upper limit on the MIR
background at $6\mu$m is $\sim10^4{\rm Jy~sr^{-1}}$.  In
Figure~\ref{fig:counts2} below, we show as the dashed curve the ratio
of the flux from all sources brighter than some flux $F_\nu$ to this
upper limit. The figure shows that our maximal model, which is
marginally consistent with the $1.6\mu$m NICMOS counts, is also
marginally consistent with the upper limit on the MIR background.

\section{Results and Discussion}
\label{sec:results}

In this section, we present the galaxy counts at different MIR
wavelengths, describe the properties of the faint sources such as
typical masses and redshift distributions, and discuss the confusion
limit for {\it TPF}. 

\subsection{Mid-IR Galaxy Counts}
\label{subsec:counts}

Figure~\ref{fig:counts2} shows the cumulative galaxy counts at 6$\mu$m
in our model, using the effective dust opacity prescription from
Charlot \& Fall (2000), superimposed on the dust--free spectral model
from GISSEL00 (Bruzual \& Charlot 2000).  The upper solid curve shows
all galaxies, and the lower curve shows only those beyond redshift
$z=5$.  The dashed curve shows the contribution of the sources to the
upper limit on the 6$\mu$m background, as discussed above.

\myputfigure{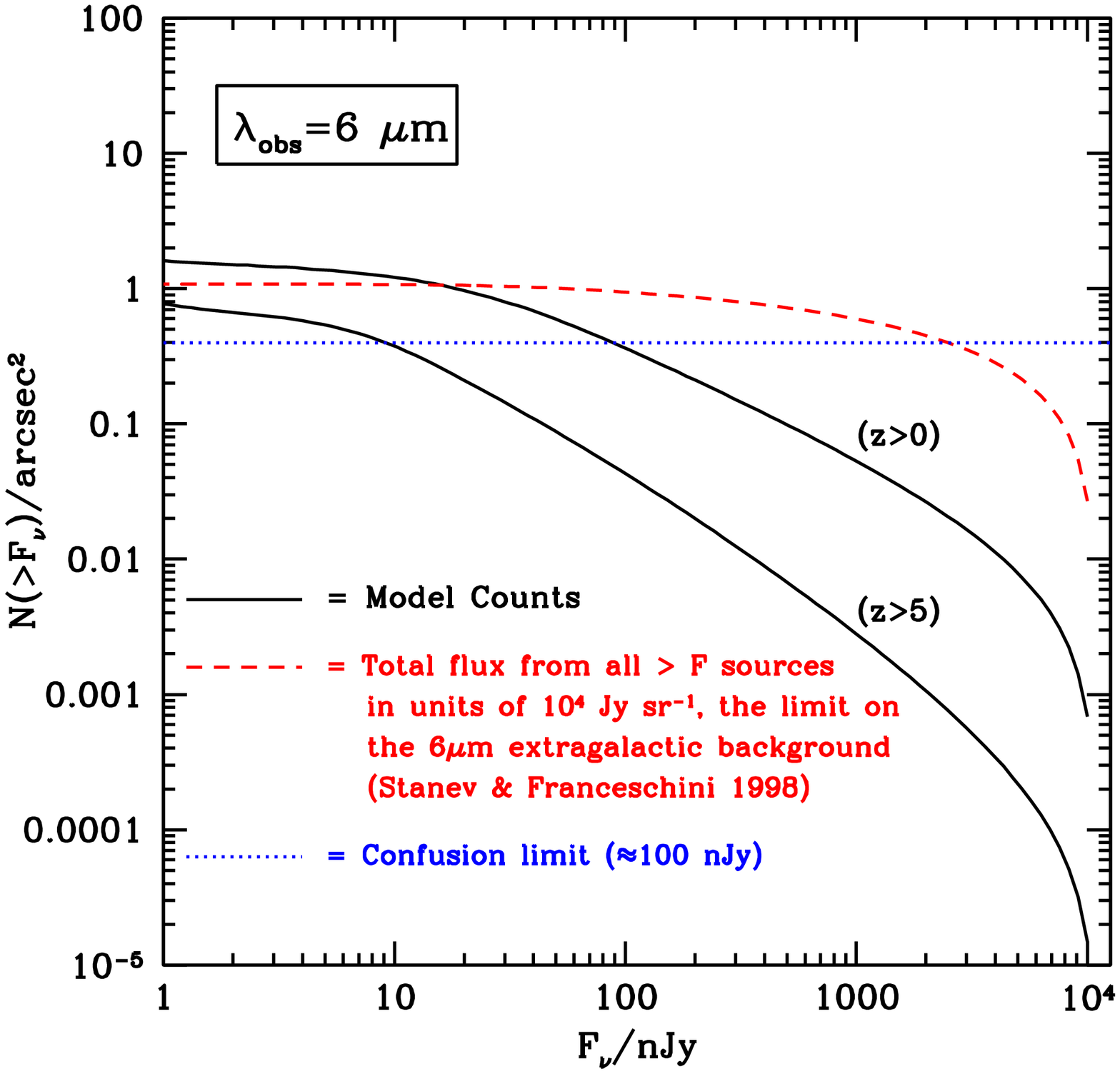}{3.2}{0.45}{-20}{-10} \figcaption{\label{fig:counts2}
Cumulative galaxy counts at 6$\mu$m. The upper solid curve shows all
galaxies, and the lower curve shows galaxies beyond redshift $z=5$.
The dashed curve shows the ratio of the total flux from all sources
brighter than the flux $F_\nu$ to the upper limit on the background,
$10^4{\rm Jy~sr^{-1}}$ (Stanev \& Franceschini 1998).  The model is
consistent with this upper limit. The dotted curve shows the confusion
limit expected from a MIR version of the TPF.}
\vspace{\baselineskip}

The most striking feature shown in Figure~\ref{fig:counts2} is the
large number of galaxies. At the flux threshold of 100 nJy, over 1000
sources are predicted per arcmin$^{-2}$.  The flattening of the counts
between $\sim$10 and $\sim$20 nJy is due to the lower limit we imposed
(\S~\ref{subsec:constraints}) on the velocity dispersion of halos that
are able to host galaxies.  It is worth emphasizing again that the
deepest data from {\it ISO} only reaches the comparatively shallow
flux threshold of $\sim0.1$mJy (at $6.7\mu$m and $15\mu$m).  In fact,
the {\it ISO} counts (at $15\mu$m) appear to show a significant
flattening from $1$mJy to $\sim0.1$mJy. This, however, is not
inconsistent with a significant re--steepening of the counts at
fainter fluxes, revealing a new population as predicted by our models.

This new population of faint, dusty, high--redshift galaxies could be
uncovered by {\it SIRTF}, although reaching the flux level of 100 nJy
for a S/N=5 detection of a point source requires an integration time
of several $\times 10^5$ seconds (Simpson \& Eisenhardt 1999).  {\it
NGST} will be able to reach the same sensitivity in a $\sim 10^4$
seconds out to $\sim10\mu$m, and in $\sim 10^6$ seconds out to
$\sim30\mu$m, while a ``NGST Deep Field'' with a $10^6$s exposure
could reach fluxes as faint as a few nJy out to $\sim10\mu$m. The flux
threshold of 100 nJy has also been chosen as the target flux for the
MIR version of {\it TPF}, based on its mission goal to detect
Earth--like planets at a distance of 10 pc.\footnote{See
http://sirtf.caltech.edu, http://www.ngst.nasa.gov, and
http://tpf.jpl.nasa.gov for quantitative discussions of the
sensitivities.}

\myputfigure{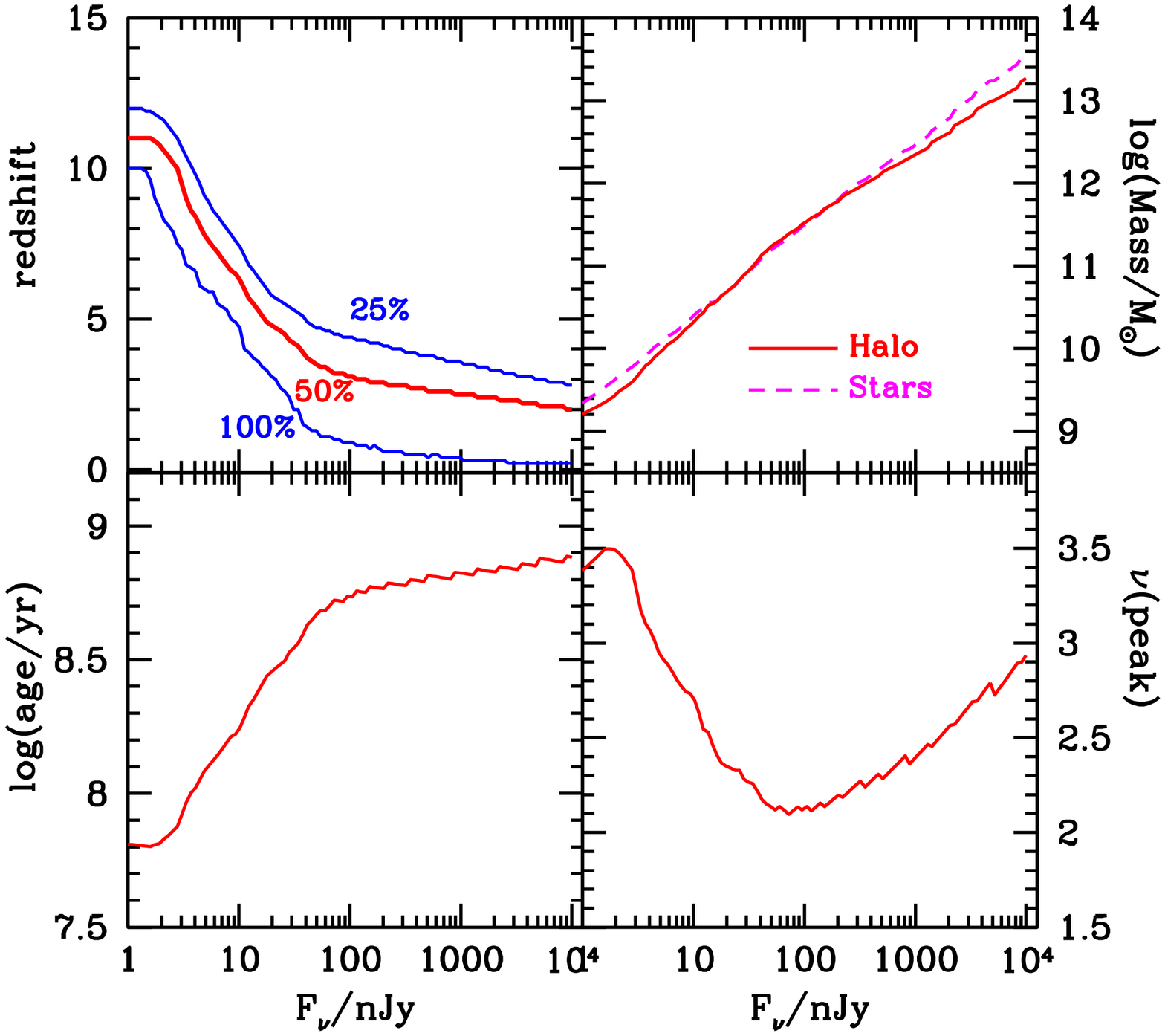}{3.35}{0.45}{-20}{-10} \figcaption{\label{fig:dndz}
Characteristic properties of the dusty sources that make up the counts
in Figures~\ref{fig:counts1} and~\ref{fig:counts2}: (a) redshifts
above which sources make up 25, 50, and 100\% of the total observed
counts; (b) masses of objects at the 50\% redshift cut; (c) typical
stellar ages; and (d) rarity of the host density peaks in units of
r.m.s. primordial density fluctuation~$\sigma_M$.}
\vspace{\baselineskip}

The characteristic properties of the sources making up the counts in
Figures~\ref{fig:counts1} and~\ref{fig:counts2} are summarized by the
four panels of Figure~\ref{fig:dndz}. In the top left panel, we
illustrate the redshift distribution of the sources as a function of
their $6\mu$m flux, by showing the redshifts at each flux beyond which
sources make up a fraction 25, 50, and 100\% of the observed counts.
The low--redshift cutoff is a result of the limit we imposed on the
circular velocities of halos harboring active galaxies.  The redshift
distribution for fainter sources is clearly biased to higher
redshifts, with an apparent upturn in the typical redshift below
$\sim100$nJy.  While approximately a half of the $100$nJy sources are
located at $z>3$, all of the $1$nJy sources are at $z>10$.  In the top
right panel, we show the mass of stars that have been converted to
stars in the halos at the 50\% redshift cut, together with the masses
of their host halos.  This explicitly demonstrates the high
starformation rates in our models, using up a nominal amount of gas
nearly equal to the halo mass -- implying multiple generations of
starformation.  The bottom left panel shows the ages of the sources at
the 50\% redshift cut.  These are between $10^8-10^9$yr, with the
fainter sources systematically younger.  Finally, the bottom right
panel shows the rarity of the density peaks hosting the halos at the
50\% redshift cut, in units of the r.m.s. primordial density
fluctuation $\sigma_M$.  The typical sources correspond to
2--3$\sigma$ peaks in the primordial density field.

\myputfigure{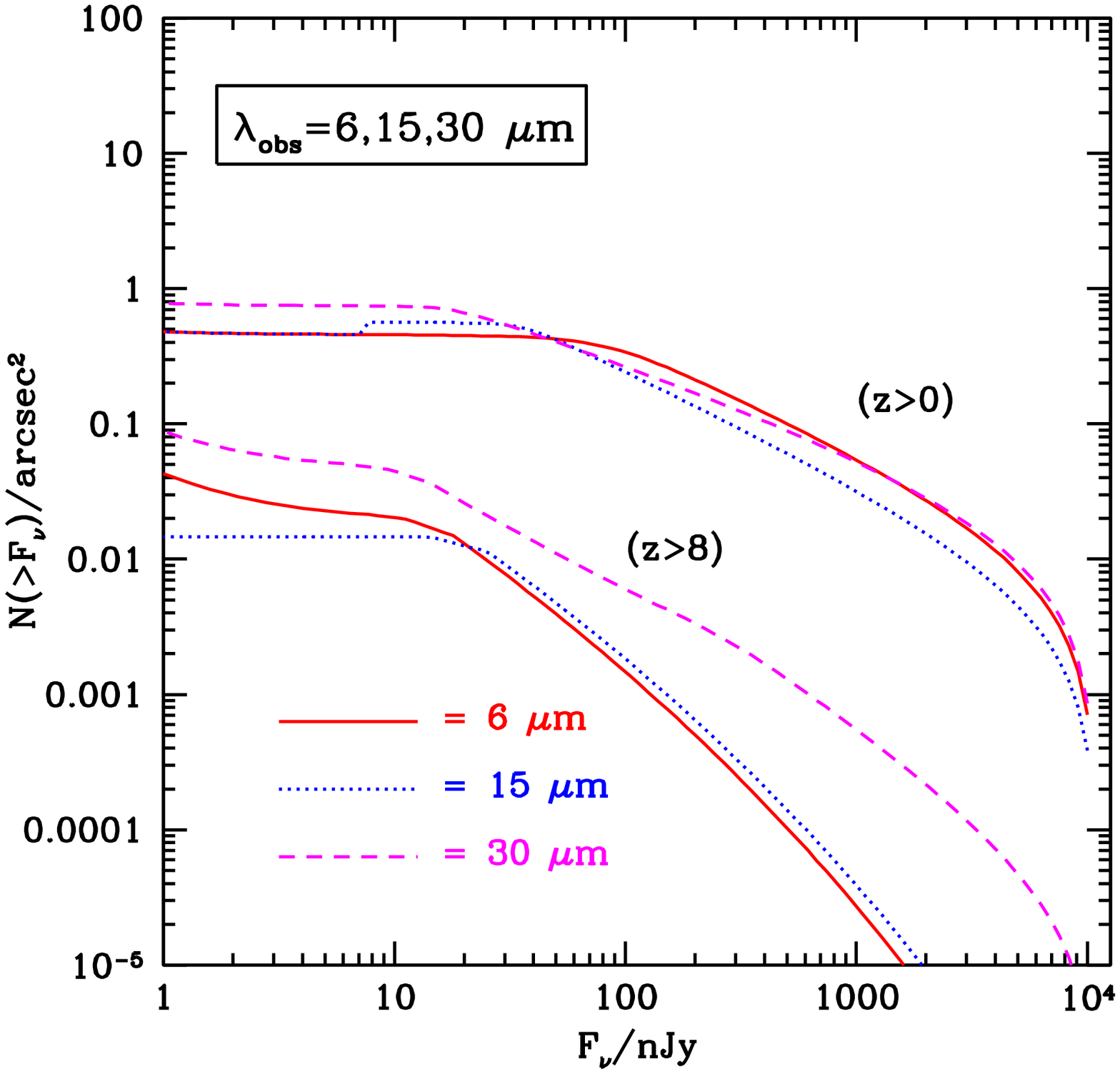}{3.2}{0.45}{-20}{-10} \figcaption{\label{fig:counts3}
Source counts as in Figure~\ref{fig:counts2} but including longer MIR
wavelengths.  To compute these counts, we have used the program GRASIL
(Silva et al. 1998) for the spectra of dusty starbursts, which
includes re--emission of starlight by hot dust.  The model is
normalized to predict the same $6\mu$m counts as in the simplified
model in Figure~\ref{fig:counts2}.  At higher redshifts, longer MIR
wavelengths are advantageous, and can reveal an increasingly larger
number of galaxies.}
\vspace{\baselineskip}

In Figure~\ref{fig:counts3} we show the cumulative galaxy counts at
longer wavelengths, using the dust models computed with the GRASIL
code.  We have re--normalized the starformation efficiency in the
GRASIL--based models so that they predict essentially the same 6$\mu$m
counts as obtained in the simplified effective dust--screen model in
Figure~\ref{fig:counts2}.  We found that this corresponds to a
reduction of the star--formation efficiency by a factor of $\sim3$
relative to equation~(\ref{eq:fstar}).  The counts at 6$\mu$m,
15$\mu$m, and 30$\mu$m are shown in Figure~\ref{fig:counts3} for all
sources (upper set of three curves), and for sources located beyond
redshift $z=8$ (lower set of three curves).  Although the 15$\mu$m
counts are somewhat below those at 6$\mu$m, the 30$\mu$m and 6$\mu$m
counts are comparable.  This follows directly from the dip in the
spectra near 15$\mu$m (see Fig.~\ref{fig:grasil}).  For the
highest--redshift sources, the advantage of going to longer
wavelengths is increased, with nearly an order of magnitude more
sources at the 100nJy threshold at 30$\mu$m than at $6\mu$m.  A
considerable number of $z>8$ galaxies, $\sim30~{\rm arcmin^{-2}}$,
are detectable at 30$\mu$m at $100$nJy.

\subsection{Confusion Noise}
\label{subsec:confusion}

The potentially large number of detectable sources raises the important
question of confusion.  For an instrument whose angular resolution elements
have an effective solid angle $\Delta\Omega$, one can define the confusion
limit such that a source at this flux corresponds to (say) a $3\sigma$
fluctuation of the unresolved background due to all fainter sources.  The
critical surface density of background sources according to this definition
depends on the slope of the counts (see, e.g., equation 8.26 in Franceschini
2000).  The slope we find in the flux range $\sim 10 - 10^4$ nJy is close to
$d\log N/d\log F \approx -1$ (see Figs~\ref{fig:counts2} and
\ref{fig:counts3}), implying that confusion limits sets in at the surface
density of 1 source per $\sim9$ beams.

For the MIR version of {\it TPF}, the effective beam--size is 0.25 arcsec$^2$,
and hence this instrument would be confusion limited at the source surface
density of $\sim0.4$ arcsec$^{-2}$. This limit is shown as the dotted line in
Figure~\ref{fig:counts2}. Although {\it TPF} is an interferometer with high
resolution and exquisite nulling, the beam--size reflects the total collecting
area of the side--lobes, and is relatively large
\footnote{See the TPF Handbook at
http://tpf.jpl.nasa.gov/library/tpf\_book/index.html.}  

For reference, we note that the effective size for the resolution element on
{\it SIRTF} at $\sim 8\mu$m is $\sim$ 1.4 arcsec$^2$ (Simpson \& Eisenhardt
1999), and our models would predict a confusion limit of $\sim 1\mu$Jy (in
rough agreement, though somewhat higher, than the estimate by Simpson \&
Eisenhardt 1999 of $0.5\mu$Jy, based on the extrapolated model number counts of
Franceschini et al. 1991).  For {\it NGST}, the size of the resolution element
in the wavelength range $\sim1-3.5\mu$m is much smaller, 0.0025 arcsec$^2$ (see
Gillett \& Mountain 1998), and the predicted counts at shorter wavelengths are
also somewhat lower than at Mid-IR; based on Figure~\ref{fig:counts1}, we do
not expect source confusion to be a problem down to 1nJy.

As can be seen from
Figures~\ref{fig:counts2} and~\ref{fig:counts3}, this surface density is
reached near $\sim 100$nJy, i.e. close to the requisite target flux to detect
Earth--like planets at 10 pc.  High redshift galaxies can cause other problems
for {\it TPF}.  The surface density of galaxies at the confusion limit implies
that in 1 out of 10 pointings, a galaxy with $F_\nu\sim100$nJy may be detected
within the {\it TPF} beam.  Every detected planet can therefore have upto 10\%
chance of being a mis--identified galaxy.  Whether this is a significant
contamination will, of course, depend on the rate at which planets are
discovered by {\it TPF}. The unresolved background could also require greater
uv--plane coverage to obtain an unambiguous image.  We emphasize that none of
these problems are likely to be show--stoppers for {\it TPF}: even for the
maximum allowed surface density, repeat observations can be used to eliminate
confusion--related problems.

\section{Conclusions}
\label{sec:conclusions}

Deep mid--infrared observations of the universe could reveal a new
population of ultra--faint, high--redshift ($z>3$) dusty starburst
galaxies that are the progenitors of present--day spheroids or bulges.
Although at a flux level of $\sim 100$nJy these sources are beyond the
reach of current instruments, the new population could be uncovered by
{\it SIRTF} and {\it NGST}, and it could also constitute background
noise for {\it TPF}.  We used a simplified semi--analytic galaxy
formation scheme to quantify the mid--infrared galaxy counts in an
extreme model, designed to maximize the number of detectable sources
down to a few nJy, while being consistent with various existing
constraints.  The model incorporates the formation of heavily
dust--enshrouded stellar populations at high redshift.

Our results show that a new population could turn up at a flux level
of $\sim 100$nJy. The sources would have typical halo masses of
$\sim10^{11}~{\rm M_\odot}$ (corresponding to $\sim 2\sigma$ peaks of
the density field), redshifts $z>3$ (with a significant tail at
$z\gsim 8$), and ages $\sim3\times10^8$yr.  The models predict upto
0.4 galaxies/arcsec$^2$ at the threshold of 100 nJy at 6$\mu$m, with a
comparable or larger surface densities at longer MIR wavelengths,
especially at the highest redshifts.  These results indicate that 
high--redshift galaxies could potentially necessitate repeat
observations with {\it TPF}.  However, the discovery of these faint
sources would be a unique and direct probe of the earliest galaxies,
and a combination of several wavelengths should provide insight into
their formation mechanism. 

\acknowledgements

We thank Gian Luigi Granato and Laura Silva for help with the use of their
program GRASIL, and Rodger Thompson for helpful commentary on the NICMOS
data. We also thank Peng Oh and Gian Luigi Granato for useful comments, and the
Princeton TPF group for stimulating discussions. ZH was supported by NASA
through the Hubble Fellowship grant HF-01119.01-99A, awarded by the Space
Telescope Science Institute, which is operated by the Association of
Universities for Research in Astronomy, Inc., for NASA under contract NAS
5-26555.  ELT was supported by the NSF grant AST98-02802 and DNS was supported
by the NASA grant NAG5-7154.


\begin{references}

\reference{} Barkana, R., \& Loeb, A. 2001, Physics Reports, in press, astro-ph/0010468

\reference{} Binney, J., \& Tremaine, S. 1987, Galactic Dynamics, Princeton Univ. Press, Princeton, NJ

\reference{} Bruzual, G., \& Charlot, S. 2000, unpublished.  The models are available from the anonymous ftp site ftp.ias.edu.

\reference{} Charlot, S., \& Fall, S. M. 2000, ApJ, 539, 718

\reference{} Clements, D. L., Desert, F-X., Franceschini, A., Reach, W. T., Baker, A. C., Davies, J. K., \& Cesarsky, C. J. 1999, A\&A, 346, 383

\reference{} Devriendt, J. E. G., \& Guiderdoni, B. 2000, A\&A, 363, 851

\reference{} Devriendt, J. E. G., Guiderdoni, B., \& Sadat, R. 1999, A\&A, 350, 381

\reference{} Dey, A., Spinrad, H., Stern, D., Graham, J. R., Chaffee, F. 1998, ApJ, 498, L93

\reference{} Draine, B. T., \& Lee, H. M. 1984, ApJ, 285, 89

\reference{} Dwek, E. 2001, in ``The Extragalactic Infrared Background and its Cosmological Implications'', proc. of IAU Symp. 204, eds. M. Harwit and M. G. Hauser, in press, preprint astro-ph/01005363

\reference{} Eisenstein, D. J., \& Hu, W. 1999, ApJ, 511, 5

\reference{} Fan, X., et al. 1999, AJ, 118, 1

\reference{} Fan, X., et al. 2000, AJ, 119, 1

\reference{} Fan, X., et al. 2001, in preparation

\reference{} Franceschini, A., Toffolatti, L., Mazzei, P., Danese, L., \& De Zotti, G. 1991, A\&ASS, 89, 285

\reference{} Franceschini, A. 2000, in ``Galaxies at High Redshift'', proc. of the XI. Canary Island Winter School of Astrophysics, eds. F. Sanchez, I. Perez-Fournon, M. Balcells, F. Moreno-Insertis, Cambridge University Press

\reference{} Franceschini, A., Aussel, H., Bressan, A., Cesarsky, C. J., Danese, L., De Zotti, G., Elbaz, D., Granato, G. L., Mazzei, P., \& Silva, L. 1997, preprint astro-ph/9707080

\reference{} Gillett, F. C., \& Mountain, M. 1998, in Science with the NGST, ASP Conference Series Vol. 133, ed. E. P. Smith \& A. Koratkar, p. 42

\reference{} Gordon, K. D., Calzetti, D., \& Witt, A. N. 1997, ApJ, 487, 625

\reference{} Guiderdoni, B., Hivon, E., Bouchet, F. R., \& Maffei, B. 1998, MNRAS, 295, 877

\reference{} Haiman, Z., Abel, T., \& Rees, M. J. 2000, ApJ, 534, 11

\reference{} Haiman, Z., \& Loeb, A. 1997, ApJ, 483, 21

\reference{} Haiman, Z., \& Loeb, A. 1998, ApJ, 503, 505

\reference{} Haiman, Z., \& Spaans, M. 1999, ApJ, 518, 138

\reference{} Haiman, Z., Spaans, M., \& Quataert, E. 2000, ApJL, 537, 5

\reference{} Haiman, Z., Thoul, A., \& Loeb, A. 1996, ApJ, 464, 523 (HTL96)

\reference{} Hu, E. M., McMahon, R. G., \& Cowie, L. L. 1999, ApJ, 522, L9

\reference{} Jenkins, A., Frenk, C. S., White, S. D. M., Colberg, J. M., Cole, S., Evrard, A. E., Couchman, H. M. P., Yoshida, N. 2001, MNRAS, 321, 372

\reference{} Kauffmann, G., \& White, S. D. M. 1993, MNRAS, 261, 921

\reference{} Lacey, C. G., \& Cole, S. 1993, MNRAS, 262, 627

\reference{} Leitherer, C., et al. 1999, ApJS, 123, 3

\reference{} Navarro, J. F., Frenk, C. S., \& White, S. D. M. 1997, ApJ, 490, 493

\reference{} Navarro, J. F., \& Steinmetz, M. 1997, ApJ, 478, 13

\reference{} Oh, S. P., \& Haiman, Z., \& Rees, M. J. 2001, ApJ, 553, 73

\reference{} Press, W. H., \& Schechter, P. L. 1974, ApJ, 181, 425

\reference{} Puget, J. L., L\'eger, A., \& Boulanger, F. 1985, A\&A, 142, L19

\reference{} Wechsler, R. H., Somerville, R. S., Bullock, J. S., Kolatt, T. S., Primack, J. R., Blumenthal, G. R., Dekel, A. 2001, ApJ, 554, 85

\reference{} Serabyn, G. et al. 1999, ``A Mid-Infrared Camera for the Next Generation Space Telescope'', a report to NASA available at ${\rm http://ngst.gsfc.nasa.gov/public\_docs.html}$

\reference{} Sheth, R., Mo, H. J., \& Tormen, G. 2001, ApJ, 323, 1

\reference{} Silva, L., Granato, G. L., Bressan, A., \& Danese, L. 1998, ApJ, 509, 103

\reference{} Silva, L., Granato, G. L., Lacey, C., \& Baugh, C. M. 1999, in ``Evolution of Galaxies on Cosmic Timescales'', proceedings of conference held at Puerto de la Cruz, Spain, December 1998, preprint astro-ph/9903350

\reference{} Simpson, C., \& Eisenhardt, P. 1999, PASP, 111, 691

\reference{} Spinrad, H., Stern, D., Bunker, A. J., Dey, A., Lanzetta, K., Yahil, A., Pascarelle, S., \& Fern\'andez-Soto, A. 1998, AJ, 117, 2617

\reference{} Stern, D., Spinrad, H., Eisenhardt, P., Bunker, A. J., Dawson, S., Stanford, S., \& Elston, R. 2000, ApJ, 533, L75

\reference{} Stanev, T., \& Franceschini, A. 1998, ApJ, 494, L159

\reference{} Thompson, R. I., Storrie-Lombardi, L. J., Weymann, R. J., Rieke, M. J., Schneider, G., Stobie, E., \& Lytle, D. 1999, ApJ, 117, 17

\reference{} Tumlinson, J., \& Shull, J. M. 2000, ApJ, 528, L65

\reference{} Weymann, R. J., Stern, D., Bunker, A., Spinrad, H., Chaffee, F. H., Thompson, R. I., \& Storrie-Lombardi, L. J. 1998, ApJ, 505, L95

\reference{} Whitmore, B. C., Kirshner, R. P., \& Schechter, P. L. 1979, ApJ, 234, 68

\reference{} Williams, R. E., et al. 1996, AJ, 112, 1335

\reference{} Zheng, W., et al. 2000, AJ, 120, 1607

\end{references}
\end{document}